\documentstyle[floats,aps,aipbook,psfig]{revtex}
\begin{document}
\title{Likelihood Methods and Classical Burster Repetition}
\author{Carlo Graziani}
\address{NRC/NASA Goddard Space Flight Center}
\author{Donald Q. Lamb}
\address{Department of Astronomy and Astrophysics, University of Chicago}
\maketitle
\begin{abstract}
We develop a likelihood methodology which can be used to search for
evidence of burst repetition in the BATSE catalog, and to study the
properties of the repetition signal.  We use a simplified model of
burst repetition in which a number $N_{\rm r}$ of sources which
repeat a fixed number of times $N_{\rm rep}$ are superposed upon a
number $N_{\rm nr}$ of non-repeating sources.  The instrument exposure
is explicitly taken into account.  By computing the likelihood for the
data, we construct a probability distribution in parameter space that
may be used to infer the probability that a repetition signal is
present, and to estimate the values of the repetition parameters.  The
likelihood function contains contributions from all the bursts,
irrespective of the size of their positional errors --- the more uncertain
a burst's position is, the less constraining is its contribution.  Thus
this approach makes maximal use of the data, and avoids the ambiguities
of sample selection associated with data cuts on error circle size.
We present the results of tests of the technique using synthetic data
sets.
\end{abstract}

\section*{Introduction}

Classical gamma-ray burst repetition is one of the most startling
possibilities suggested by analyses the BATSE 1B catalog.  Quashnock \&
Lamb \cite{ql93} searched for evidence of strong clustering of burst
positions by studying the distribution of nearest-neighbor separations,
and reported an excess of bursts clustered on 5$^\circ$ scales with a
significance of $2.5\times 10^{-4}$.  Wang \& Lingenfelter
\cite{wl94,wl95} have studied clustering of bursts in the 1B catalog in
both angle and time, and found clustering on scales of 4 days and
4$^\circ$ with a significance of $2\times 10^{-5}$.  Quashnock
\cite{q95} finds the odds favoring repetition over no repetition to be
about 4:1, based on a counts-in-cells likelihood technique.

On the other hand, the data in the BATSE 2B $-$ 1B catalog (that is,
those bursts seen in the 2B catalog after the end of the 1B epoch) show
no such significant repetition signal \cite{m95}.  This is not
unexpected, given the lower sky exposure due to the failure of the tape
recorders aboard CGRO.  Nevertheless, the absence of confirmation of
the 1B signals, combined with the only moderate significance of those
signals, have fueled a controversy over whether the evidence for burst
repetition in the 1B catalog is merely a statistical fluctuation,
perhaps amplified by choice of bin intervals (in the case of the
time-angle analysis) or of cuts in error circle size (in the case of
the nearest-neighbor analysis).  These criticisms have been answered
convincingly through analysis of the effect of data cuts \cite{ql93}
and of choice of binsize \cite{wl95}.  Nevertheless the evidence for
repetition is not universally regarded as conclusive.

Under the circumstances, there is a premium on the development of
techniques to probe for burst repetition that are more sensitive
than existing ones --- to achieve higher significance detections ---
and that do not rely on either binning or data cuts.  In this work
we present such a technique, founded upon the likelihood function for
a simple model of burst repetition.  We also exhibit the results of
tests of the method on synthetic data sets, which allow us to gauge
its sensitivity and to understand potential systematic effects.

\section*{The Likelihood Function}

In order to calculate the likelihood function, we require a model of
burst repetition.  The model we have adopted is one in which $N_{\rm
r}$ repeating sources are superposed upon $N_{\rm nr}$ non-repeating
sources.  Each repeating source bursts exactly $N_{\rm rep}$ times,
although not all these bursts are observed because of the limited sky
exposure.  This simplified model captures the essential features of the
phenomenon --- number of repeating sources and repetition rate ---
while permitting practical calculation.  The model parameters are
$N_{\rm r}$ and $N_{\rm rep}$.  $N_{\rm nr}$ is set by the data when
$N_{\rm r}$ and $N_{\rm rep}$ are chosen, and the sky exposure $E$ is
fixed at the value appropriate for the epoch of the BATSE catalog under
study.

The likelihood function calculation requires that we identify
likely ``partitions'' $\pi$ of the data.  A partition is an assignment
of some bursts to each of the repeating sources.  The likelihood
function $L(N_{\rm r},N_{\rm rep})$ is then given by
\begin{eqnarray}
L(N_{\rm r},N_{\rm rep}) &=& \sum_\pi (N_{\rm r}+N_{\rm nr})!\nonumber\\
&&\times\left\{\prod_{j=1}^{N_{\rm r}}\left[ {N_{\rm rep}!\over n_j!(N_{\rm
rep}-n_j)!} E^{n_j}(1-E)^{N_{\rm rep}-n_j}\times {\cal L}_j\right]\right\},
\label{likefn}
\end{eqnarray}
where $n_j$ is the observed number of bursts from the $j$th source
(according to the partition $\pi$), $E$ is the sky exposure, and ${\cal
L}_j$ is the ``cluster likelihood'' for the $j$th source of $\pi$.
${\cal L}_j$ acts as a measure of the plausibility of the assignment of
the $n_j$ bursts to the source, and is in fact the expression
for the Bayesian odds favoring the hypothesis that the $n_j$ bursts
were produced by the same source over the hypothesis that they each had
a separate source.  Assuming that the scatter of the observed positions
about the true source positions is described by the Fisher
distribution, ${\cal L}_j$ is given by the following expression:
\begin{equation}
{\cal L}_j = {{\Sigma_j}^2\sinh{{\Sigma_j}^{-2}}\over
\prod_{k=1}^{n_j}\left[{\sigma_{jk}}^2\sinh{{\sigma_{jk}}^{-2}}\right]}
\quad ;\quad
\Sigma_j \equiv
\left\|\sum_{k=1}^{n_j}{\vec{x}_{jk}\over{\sigma_{jk}}^2}\right\|^{-1/2}.
\label{clik_f}
\end{equation}
Here, $\vec{x}_{jk}$ is the unit vector describing the position on the sky
of the $k$th burst of the $j$th source, and $\sigma_{jk}$ is the error
in that position.  $\Sigma_j$ is the error in the maximum likelihood
position of the $j$th source.  The maximum likelihood position itself
is $\vec{z}_j={\Sigma_j}^2\sum_{k=1}^{n_j}\vec{x}_{jk}/{\sigma_{jk}}^2$.

The meaning of Eq.\ (\ref{clik_f}) may be clarified by taking the
small-angle limit, in which all the $\sigma$ are much less than one.
In this limit the Fisher distribution becomes the two-dimensional,
symmetric Gaussian distribution, and the expression for the $j$th cluster
likelihood becomes
\begin{equation}
{\cal L}_j \approx {{1\over 2}{\Sigma_j}^2
\over
\prod_{k=1}^{n_j}\left({{\sigma_{jk}}^2\over 2}\right)}
\times
\exp\left[-{1\over 2}\sum_{k=1}^{n_j}
{(\vec{x}_{jk}-\vec{z}_j)^2\over{\sigma_{jk}}^2}\right]
\quad ;\quad
\Sigma_j \approx \left[\sum_{k=1}^{n_j}{1\over {\sigma_{jk}}^2}\right]^{-1/2}.
\label{clik_g}
\end{equation}

From Eq.\ (\ref{clik_g}) we see that ${\cal L}_j$ is a product of two
factors:  the exponential factor, which penalizes (decreases) the
likelihood if the $j$th cluster of $\pi$ is too dispersed, and the
rational factor, which rewards (increases) the likelihood for making $n_j$
as large as possible (recall that $\sigma_{jk}\ll 1$).  These two factors
compete with each other, since the addition of a burst to a cluster may well
provoke a harsher penalty from the exponential factor (if the burst is
implausibly distant from the rest of the cluster) than the reward it
reaps from the rational factor.  This fact allows ${\cal L}_j$ to be used
as a tool in identifying ``optimum'' clusters of bursts.  This is an
important observation, since the sum over partitions $\pi$ in the
likelihood clearly ranges over an impossibly large number of
configurations.  Fortunately, most partitions contribute negligibly to
the likelihood, as they assign bursts to improbably dispersed
clusters.  We may thus restrict our attention to partitions that make
appreciable contributions to the likelihood by stringing together into
partitions only ``optimum'' clusters identified using ${\cal L}_j$.
This is a key ingredient of our technique.

Note that the dependence of ${\cal L}_j$ on the burst positions is
strongest for bursts with small error circles and weakest for bursts
with large error circles.  This is the reason that the inclusion of
bursts with poorly determined positions does not weaken the repetition
signal --- such bursts are individually unable to seriously constrain
inferences based on the likelihood.  On the other hand, even bursts
with large error circles bear {\it some} information, particularly if
there are many of them.  Our approach allows that information to be
exploited to the fullest extent possible.

We use Bayes theorem to interpret the likelihood function as a
probability distribution in $(N_{\rm r},N_{\rm rep})$ parameter space.
This distribution is the basis for all our statistical inferences.  By
comparing the contribution of the likelihood function to the entire
parameter space to the value of the likelihood for the non-repeating
model ($N_{\rm r}=0$), we may compute the odds favoring the repetition
hypothesis over the no-repetition hypothesis, thus obtaining a
sensitive test for the presence of a repetition signal.  Furthermore,
the detailed shape of the distribution allows us to estimate the
maximum likelihood parameter values, and to infer credible regions for
those parameters.

\section*{Testing the Technique}

We have tested the method on synthetic data, simulating both repeating
and non-repeating ($N_{\rm r}=0$) models, in order to test its
sensitivity and power, as well as to gauge the reliability of the
parameter estimates that it produces.

\begin{figure}
\psfig{figure=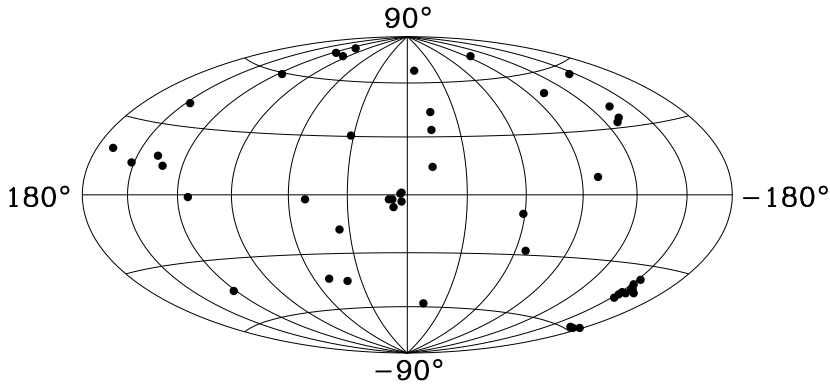,angle=90,width=\hsize,bbllx=145bp,bblly=54bp,bburx=465bp,bbury=715bp}
\caption{Sky map resulting from a simulation with $N_{\rm r}=5$,
$N_{\rm rep}=15$, and a sky exposure $E=0.33$.  Note that more than three
bursts are observed from only two of the five repeating sources, even though
$N_{\rm rep}=15$, because the sky exposure is limited.}
\label{skymap}
\end{figure}

Fig.\ \ref{skymap} shows a sky map resulting from a simulation assuming
$N_{\rm r}=5$, $N_{\rm rep}=15$, and sky exposure
$E=0.33$.  There are 25 bursts from the 5 repeating sources, and 50
bursts from all sources.  The assumed total error $\sigma_{\rm circ}$
--- that is, the radius of the 68\% circular confidence region including
both systematic and statistical error --- is 5$^\circ$ for all bursts.
$\sigma_{\rm circ}$ is related to the Fisher distribution parameters
$\sigma_{jk}$ above by ${\sigma_{jk}}^2=0.871(1-\cos\sigma_{\rm circ})$.

\begin{figure}
\psfig{figure=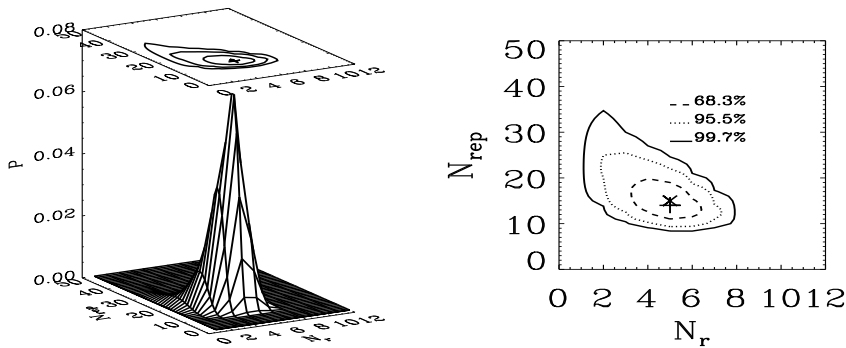,angle=90,width=\hsize,height=2.0in,bbllx=60bp,bblly=0bp,bburx=520bp,bbury=792bp}
\caption{Left panel: Probability distribution in the
($N_r$,$N_{rep}$)-plane that results from the analysis of 
the simulated data shown in
Fig.\ \ref{skymap}.  Right panel: 1-, 2-, and 3-$\sigma$ credible
regions in the ($N_r$,$N_{rep}$)-plane.  The cross shows the maximum
likelihood parameter estimates, while the X marks the true parameter
values.}
\label{likeplot}
\end{figure}

Fig.\ \ref{likeplot} shows the probability distribution in the
($N_r$,$N_{rep}$)-plane that results from the analysis of this
simulated data set.  The distribution is sharply peaked near the true
parameter values.  Also shown are the 1-, 2-, and 3-$\sigma$ credible
regions for the model parameters.

An interesting question that arises is whether the credible regions
shown in Fig.\ \ref{likeplot} are well-calibrated, in the sense that
they bracket the true parameters with frequency equal to the stated
probability contained within the contour.  To begin addressing this
question, we analyzed 20 simulated data sets with the same parameter
values as those given above.  We found that the 68\% contour included
the true parameters in 18 out of 20 simulations, which by the binomial
distribution is quite consistent with the 68\% interpretation of
the contour.  We are carrying out analyses on many more simulations
in order to refine the contour calibration.

The odds may be computed by ascribing equal a priori probability to the
repeating and the non-repeating hypothesis, and, within the repeating
hypothesis, equal a priori probability to each of the discrete
parameter values $(N_{\rm r},N_{\rm rep})$ at which the likelihood is
evaluated.  This last assignment dilutes the value of the odds if the
parameter space is very large, and accounts naturally for the ``number
of attempts'' correction to the statistical significance of the detection.

We calculated the odds for each of the 20 simulated data sets discussed
above, and found that the mean and standard deviation log odds were
$\log {\cal O}=22.5\pm 7.9$.  We also simulated 20 data sets assuming
no repetition, and calculated the odds for each of those.  The result
was $\log {\cal O}=0.5\pm 2.0$.  Clearly, then, the likelihood function
approach is quite capable of distinguishing between no repetition and
the kind of repetition simulated here, and thus offers a promise of a
sensitive and powerful statistic for the detection of repetition
signals in burst data.


\begin{references}
\bibitem{ql93} J.~M.~Quashnock, and D.~Q.~Lamb, Mon. Not. R. Astron.
Soc., {\bf 265}, L59 (1993).
\bibitem{wl94} V.~C.~Wang, and R.~E.~Lingenfelter, in {\it Gamma-Ray Bursts},
Proceedings of the Second Workshop, Huntsville, edited by G.~J.~Fishman,
J.~J.~Brainerd, and K.~Hurley (AIP, 1994), p. 160.
\bibitem{wl95} V.~C.~Wang, and R.~E.~Lingenfelter, Astrophys.~J.,
{\bf 441}, 747 (1995).
\bibitem{q95} J.~M.~Quashnock, (these proceedings).
\bibitem{m95} C.~Meegan {\it et al.}, Astrophys.~J.~Lett., in press (1995).
\end{references}
\end{document}